\documentclass[aps,twocolumn,pra,showpacs,floatfix]{revtex4}

\usepackage{graphicx}

\begin{document}
\title{Search for variation of the fundamental constants in atomic,
molecular and nuclear spectra}
\author{V. V. Flambaum and V. A. Dzuba}
\address{School of Physics, University of New South Wales, Sydney 2052,
Australia}
%\correspond{flambaum@phys.unsw.edu.au}
%\author{V. A. Dzuba}

\date{\today}

\begin{abstract}

The search for variation of the fundamental constants such as the fine
structure constant $\alpha$ ($\alpha=e^2/\hbar c$) and the ratios of
fundamental masses (e.g., electron to proton mass ratio $\mu=m_e/m_p$)
is reviewed. Strong emphasis is given to establishing the relationships 
between the change in the measured frequencies of atomic, molecular or
nuclear transitions and the corresponding change of the fundamental constants.
Transitions in which the sensitivity of the frequency change to the variation 
of the fine structure constant is strongly enhanced are discussed and
most recent experimental results are presented. Most attention is given
to the use of atomic, molecular and nuclear transitions in the study
of quasar absorption spectra and in atomic clock experiments. 

\end{abstract}

\maketitle

\section{Introduction}

Search for variation of the fundamental constants is now a very popular area
of research. It is motivated by theories unifying gravity with other interactions 
as well as many cosmological models which suggest a possibility for fundamental
constants to very in space and time. Hints that some fundamental constants 
might had slightly different values in distant past have been found
in Big Bang nucleosynthesis, quasar absorption spectra and Oklo natural
nuclear reactor (see, e.g reviews~\cite{Uzan,Flambaum07a}). 
However, most of publications report only 
constrains on possible variation of the fundamental constant.
For example, strong limits on present-day time variation of the
fine structure constant and lepton to hadron mass ratio come from
atomic clock experiments (see, e.g.~\cite{Flambaum07a,Lea}).

Good progress in the search for the manifestation of the variation 
of the fine structure constant in quasar absorption spectra was
achieved with a very sensitive many-multiplet (MM) method which was
suggested in Ref.~\cite{MM}. The analysis of the data from 143 
absorption systems revealed anomalies in atomic spectra which can
be best explained by assuming a slightly smaller value of the
fine structure constant about 10 billion years ago~\cite{murphy1}.

The MM method relies on atomic calculations to relate the
variation of the frequencies of atomic transitions to the
variation of the fine structure constant. Similar calculations
for the atomic transitions which are used or planed to be used
as atomic optical frequency standards also allowed to extend
this fast developing area of research to the study of the
present-date time variation of the fine structure and other
constants.

It only makes sense to consider variation of dimensionless
constants (see, e.g.~\cite{Uzan}). Therefore, most of the research
is focused on the study of either variation of the fine structure
constant $\alpha$ ($\alpha=e^2/\hbar c$) or electron to proton
mass ratio $\mu = m_e/m_p$ or some other fundamental mass ratio
like, e.g. \\
$m_{e,q}/\Lambda_{QCD}$ where $m_q$ is the quark mass and
$\Lambda_{QCD}$ is the quantum chromodynamics mass scale defined as
a position of Landau pole in the logarithm for the running strong
coupling constants. Models of the grand unification suggest that
if fundamental constants vary than fundamental mass ratio varies faster
that the fine structure constant and therefore it might be easier to
detect it. Transitions sensitive to the variation of the mass ratio
can be found in atomic microwave, molecular and nuclear spectra. 
Corresponding studies use quasar absorption spectra and laboratory
measurements~\cite{Flambaum07a}.

Further advance in sensitivity can be achieved if strong enhancement 
is found. The relative rate of frequency change 
($\dot{\omega}/\omega$)
due to change of the fundamental
constants is usually inversely proportional to the energy interval between
the two states $\hbar \omega$. Therefore the relative enhancement is likely to be 
found for the pairs of 
almost degenerate long-living states of significantly different
nature. Different nature of states is important to make sure that they 
don't move with the same rate. Also, long lifetime is needed for high
accuracy of the frequency measurements. Pairs of states which satisfy
all these conditions can be found in atomic, molecular and nuclear spectra.

In present paper we review the search for variation of the fundamental
constants focusing of the connections between measured values and the
change of the fundamental constants, enhancement mechanisms and most
recent experimental results.

\section{Optical atomic spectra}

The use of optical atomic transitions in the search for variation
of the fine structure constant $\alpha$ is based on the fact that the
frequencies of these transitions depend differently on $\alpha$ and
one can compare the frequencies of different transitions over long
period of time. If $\alpha$ is changing the ratio of the frequencies
would change too.

In atomic units $\alpha=1/c$ where $c$ is speed of light. Therefore,
$\alpha=0$ correspond to a non-relativistic limit. Leading relativistic 
corrections are proportional to $(Z\alpha)^2$ where $Z$ is nuclear charge.
Therefore, it is convenient to present the dependence of atomic frequencies on
the fine-structure constant $\alpha$ in the vicinity of its physical
value $\alpha_0$ in the form
\begin{equation}
  \omega(x) = \omega_0 + qx,
\label{omega}
\end{equation}
where $\omega_0$ is the present laboratory value of the frequency and
$x = (\alpha/\alpha_0)^2-1$. The sensitivity coefficient $q$ is to be
found from atomic calculations. 

Note that
\begin{equation}
 q = \left .\frac{d\omega}{dx}\right|_{x=0}.
\label{qq}
\end{equation}
To calculate this derivative numerically we use
\begin{equation}
  q \approx  \frac{\omega(+\delta) - \omega(-\delta)}{2\delta}.
\label{deriv}
\end{equation}
and vary the value of $\alpha$ in the computer code.

The same formula (\ref{omega}) and similar atomic calculations are
used for the analysis of quasar absorption spectra and for 
laboratory experiments involving optical transitions. A 
review of the methods of atomic calculations and critical 
compilation of the results for the quasar absorption spectra
are presented elsewhere~\cite{DzubaFlamb}. Below we review
the latest results of the analysis and discuss the results
of calculations for the laboratory measurements.

\subsection{Comparison of quasar absorption spectra with laboratory spectra}

The analysis of the three independent samples of data containing 143
absorption  systems spread over red shift range $0.2 <z < 4.2$
gives ~\cite{murphy1}
%$\frac{ \delta \alpha}{\alpha}= (-0.543 \pm 0.116) \times 10^{-5}$~\cite{murphy1}. 
\begin{equation}
\frac{ \delta \alpha}{\alpha}= (-0.543 \pm 0.116) \times 10^{-5}.
\label{dalphaQAS}
\end{equation}
If one assumes the linear dependence
of $\alpha$ on time, the fit of the data gives 
%$d\ln{\alpha}/dt=(6.40 \pm 1.35) \times 10^{-16}$ 
\begin{equation}
%d\ln{\alpha}/dt=(6.40 \pm 1.35) \times 10^{-16}
  \delta \alpha/\alpha = (6.40 \pm 1.35) \times 10^{-16}/{\rm yr}, 
\label{dadtQS}
\end{equation}
per year over time interval of about 12 billion years.
A very extensive search for possible
systematic errors has shown that known systematic effects can not explain
the result~\cite{Webb03a}.

Recently our method and calculations~\cite{MM,Dzuba99,Dzuba01,Dzuba02}
%\cite{dzuba1999,dzubaPRL,Dy,q1,q2,q3}
were used by two other groups \cite{chand,Levshakov,Levshakov1}. 
However, they have not detected any variation of $\alpha$.
The results of \cite{chand} were questioned in Refs.~\cite{murphy2,murphy3}. 
Re-analysis of Ref.~\cite{chand}
data revealed flawed parameter estimation methods.
The authors of \cite{murphy2,murphy3} claim that the same spectral data fitted 
more accurately give  $\frac{ \delta \alpha}{\alpha}=
(-0.64 \pm 0.36) \times 10^{-5}$ (instead of  $\frac{ \delta \alpha}{\alpha}=
(-0.06 \pm 0.06) \times 10^{-5}$ in Ref.\cite{chand}). However, even this
revised result may require further revision. 

Note  that the results of 
\cite{murphy1} are based on the data from the
Keck telescope which is located in the Northern hemisphere  (Hawaii).
The results of \cite{chand,Levshakov,Levshakov1,murphy2,murphy3}
are based on the data from the different telescope (VLT) located
in the Southern hemisphere (Chile). Therefore, some difference in the results
may appear due to the spatial variation of $\alpha$. 

\subsection{Optical atomic clocks}

\begin{table}
\begin{center}
\caption{Experimental~\cite{NIST} and theoretical energies and calculated 
$q$-coefficients (cm$^{-1}$) for optical atomic clock transitions 
from the ground state to an excited metastable state.}
\begin{tabular}{l l l l r r r  }
\label{tb:opt}
$Z$ & Atom & \multicolumn{2}{c}{State} & E(exp) & E(calc) &
\multicolumn{1}{c}{q} \\
\hline
 13 & Al~II & $3s3p$ & $^{3}P^o_{0}$ & 37393  & 37328 & 146 \\
    &      & $3s3p$ & $^{3}P^o_{1}$ & 37454  & 37393 & 211 \\
    &      & $3s3p$ & $^{3}P^o_{2}$ & 37578  & 37524 & 343 \\
    &      & $3s3p$ & $^{1}P^o_{1}$ & 59852  & 60090 & 278 \\
	   
 20 & Ca~I  & $4s4p$ & $^{3}P^o_{0}$ & 15158  & 15011 & 125 \\
    &      & $4s4p$ & $^{3}P^o_{1}$ & 15210  & 15066 & 180 \\
    &      & $4s4p$ & $^{3}P^o_{2}$ & 15316  & 15179 & 294 \\
    &      & $4s4p$ & $^{1}P^o_{1}$ & 23652  & 24378 & 250 \\
	         				   						       
 38 & Sr~I  & $5s5p$ & $^{3}P^o_{0}$ & 14318  & 14169 &  443 \\
    &      & $5s5p$ & $^{3}P^o_{1}$ & 14504  & 14367 &  642 \\
    &      & $5s5p$ & $^{3}P^o_{2}$ & 14899  & 14786 & 1084 \\
    &      & $5s5p$ & $^{1}P^o_{1}$ & 21698  & 22305 &  924 \\

 38 & Sr~II & $4d$   & $^2D_{3/2}$   & 14556 & 14550 &  2828 \\
    &       & $4d$   & $^2D_{5/2}$   & 14836 & 14880 &  3172 \\
	   
 47 & Ag~I  & $4d^{9}5s^2$ & $^2D_{5/2}$ & 30242 & 30188 & -11300 \\
    &       & $4d^{9}5s^2$ & $^2D_{3/2}$ & 34714 & 35114 &  -6500  \\
 49 & In~II & $5s5p$ & $^{3}P^o_{0}$ & 42275  & 42304 & 3787 \\
    &      & $5s5p$ & $^{3}P^o_{1}$ & 43349  & 43383 & 4860 \\
    &      & $5s5p$ & $^{3}P^o_{2}$ & 45827  & 45904 & 7767 \\
    &      & $5s5p$ & $^{1}P^o_{1}$ & 63034  & 62325 & 6467 \\
	         				   						       
 56 & Ba~II & $5d$   & $^2D_{3/2}$   &  4844 &  5104 &  6104 \\
    &       & $5d$   & $^2D_{5/2}$   &  5675 &  6040 &  6910 \\

 70 & Yb~I  & $6s6p$ & $^{3}P^o_{0}$ & 17288  & 16950 & 2714 \\
    &       & $6s6p$ & $^{3}P^o_{1}$ & 17992  & 17705 & 3527 \\
    &       & $6s6p$ & $^{3}P^o_{2}$ & 19710  & 19553 & 5883 \\
    &       & $6s6p$ & $^{1}P^o_{1}$ & 25068  & 26654 & 4951 \\

 70 & Yb~II & $4f^{13}6s^2$ & $^2F^o_{7/2}$ & 21419 & 20060 & -63752  \\
    &       & $4f^{14}5d$   & $^2D_{3/2}$   & 22961 & 23942 & 11438 \\
    &       & $4f^{14}5d$   & $^2D_{5/2}$   & 24333 & 25503 & 12582 \\
    &       & $4f^{13}5d6s$ & $^3[\frac{3}{2}]^o_{5/2}$ & 26759 & 26781 & -46863  \\
    &       & $4f^{13}6s6p$ & $(\frac{7}{2},0)_{\frac{7}{2}}$ & 47912 & 47927 & -60432  \\
 70 & Yb~III &$4f^{13}5d$   & $(\frac{5}{2},\frac{5}{2})^o_0$ & 45277 & 46505 & -32800  \\

 79 & Au~I  & $5d^{9}6s^2$  & $^2D_{5/2}$ &  9161 &  9186 & -38550  \\
    &       & $5d^{9}6s^2$  & $^2D_{3/2}$ & 21435 & 22224 & -26760  \\
	   
 80 & Hg~I  & $6s6p$ & $^{3}P^o_{0}$ & 37645  & 37420 & 15299 \\
    &       & $6s6p$ & $^{3}P^o_{1}$ & 39412  & 39299 & 17584 \\
    &       & $6s6p$ & $^{3}P^o_{2}$ & 44043  & 44158 & 24908 \\
    &       & $6s6p$ & $^{1}P^o_{1}$ & 54069  & 56219 & 22789 \\

 80 & Hg~II & $5d^{9}6s^2$ & $^2D_{5/2}$ & 35515 & 35066 & -52200  \\
    &       & $5d^{9}6s^2$ & $^2D_{3/2}$ & 50556 & 50886 & -37700  \\
	   
 81 & Tl~II & $6s6p$ & $^{3}P^o_{0}$ & 49451  & 49865 & 16267  \\
    &       & $6s6p$ & $^{3}P^o_{1}$ & 53393  & 52687 & 18845  \\
    &       & $6s6p$ & $^{3}P^o_{2}$ & 61725  & 62263 & 33268  \\
    &       & $6s6p$ & $^{1}P^o_{1}$ & 75600  & 74717 & 29418  \\

 88 & Ra~II & $6d$   & $^2D_{3/2}$ & 12084 & 11882 & 18150 \\
    &       & $6d$   & $^2D_{5/2}$ & 13743 & 13593 & 19000 \\
\hline
\end{tabular}
\end{center}
\end{table}

Study of the atomic optical frequency standards is huge and fast developing
area of research. The accuracy of the measurements of atomic transitions
between the ground and a metastable excited state for some atoms are on the
level of $10^{-16}$ with good prospects for further improvements (see, e.g.~\cite{Lea}).
This makes it possible to use atomic frequency standards to study 
present-day time variation of $\alpha$  and other fundamental constants.
This is done by comparing frequencies of different atomic transition over
long period of times. One has to know how the frequencies depend of $\alpha$
for the interpretation of the results.
We have performed relativistic many-body calculations
\cite{Dzuba99,Dzuba00c,Dzuba03c,Angstmann,Dzuba05c,Karshenboim,Dzuba08c}
similar to those used in the analysis of the quasar absorption spectra
to establish this dependence.
The results are summarized in Tables~\ref{tb:opt} and ~\ref{q-s}.
Calculated energies are included to illustrate the accuracy of the
calculations.
The $q$ coefficients for optical clock transitions may be 
substantially larger than those used in cosmic transitions since the clock 
transitions are often in heavy atoms (Hg II, Yb II, Yb III, etc.) while 
cosmic spectra contain mostly  lines of light atoms ($Z <33$). The relativistic 
effects are proportional to $Z^2 \alpha^2$.

Combined analysis of the frequency shifts of optical transitions in
H~\cite{H04}, Yb$^+$~\cite{Yb06}, Hg$^+$~\cite{Fortier} and Sr~\cite{Sr08}
as measured against Cs primary frequency standard gives the following
limit on the rate of time-change of $\alpha$ and $\mu=m_e/m_p$~\cite{Sr08}
\begin{eqnarray}
%%  \delta \alpha/\alpha &=& (-3.1 \pm 3.0) \times 10^{-16}/{\rm yr}, 
  \delta \alpha/\alpha &=& (-3.3 \pm 3.0) \times 10^{-16}/{\rm yr}, 
\label{dalpha} \\
%%  \delta \mu/\mu &=& (1.5 \pm 1.7) \times 10^{-15}/{\rm yr}.
  \delta \mu/\mu &=& (1.6 \pm 1.7) \times 10^{-15}/{\rm yr}.
\label{dmu0}
\end{eqnarray}
In a more recent work~\cite{Rosenband} in which optical frequencies
of Al$^+$ and Hg$^+$ were compared over a year the variation of
$\alpha$ is limited by
\begin{equation}
  \delta \alpha/\alpha = (-1.6 \pm 2.3) \times 10^{-17}/{\rm yr}.
\label{HgAl}
\end{equation}
The results for $\delta \alpha/\alpha$ (\ref{dalpha},\ref{HgAl}) 
disagree with the result (\ref{dadtQS}) obtained from astrophysical
observation.
Note however, that eq.~(\ref{dadtQS}) assumes the same rate of change over
about 10 billion years which doesn't need to be true.

\subsection{Enhanced relative effects of $\alpha$ variation in atoms}

\begin{table*}
\begin{center}
\caption{Experimental and theoretical energies and calculated
relativistic energy shifts ($q$-coefficients, cm$^{-1}$) for
pairs of almost degenerate states of Dy and Ho.}
\label{q-s}
  \begin{tabular}{c l l l l l l r r}
 $Z$ & Atom &  \multicolumn{2}{c}{Ground state} &  \multicolumn{2}{c}{Excited state} &
\multicolumn{2}{c}{Energy[cm$^{-1}$]} &
\multicolumn{1}{c}{$q$ [cm$^{-1}$]} \\
     & Configuration & $J$ & Configuration & $J$ & Expt.\cite{NIST} & Theor. &  \\
\hline
%47 & Ag~I   & $4d^{10}5s$    & 1/2 & $4d^{9}5s^2$  & 5/2 & 30242.26~ & 30188 & -11300 \\
%   &        &                &     & $4d^{9}5s^2$  & 3/2 & 34714.16~ & 35114 &  -6500  \\
66 & Dy~I   & $4f^{10}6s^2$  &  8 & $4f^{10}5d6s$  & 10  & 19797.96~ & 20077 &   7952  \\
   &        &                &    & $4f^{9}5d^26s$ & 10  & 19797.96~ & 19693 & -25216  \\

67 & Ho~I   & $4f^{11}6s^2$  & 15/2 & $4f^{10}5d6s^2$ & 11/2 & 20493.40 & 21763 &  -28200  \\
   &        &                &      & $4f^{11}5d6s$   & 13/2 & 20493.77 & 20872 &    7300  \\
   &        &                &      & $4f^{11}6s6p$   & 13/2 & 22157.86 & 22599 &    3000  \\
   &        &                &      & $4f^{11}5d6s$   &  9/2 & 22157.88 & 22631 &    8000  \\

%70 & Yb~II  & $4f^{14}6s$    & 1/2 & $4f^{13}6s^2$ & 7/2 & 21418.75~ & 20060 & -63752  \\
%   &        &                &     & $4f^{14}5d$   & 3/2 & 22960.80~ & 23942 &  11438  \\
%   &        &                &     & $4f^{14}5d$   & 5/2 & 24332.69~ & 25503 &  12582  \\
%   &        &                &     & $4f^{13}6s^2$ & 5/2 & 31568.08~ & 31303 & -53400  \\
%   &        &                &     & $4f^{13}5d6s$ & 5/2 & 26759.02~ & 26781 & -46863  \\
%   &        &                &     & $4f^{13}6s6p$ & 7/2 & 47921.31~ & 47927 & -60432  \\
%70 & Yb~III & $4f^{14}$      &  0  & $4f^{13}5d$   &  0  & 45276.85~ & 46505 & -32800  \\
%79 & Au~I   & $5d^{10}6s$    & 1/2 & $5d^{9}6s^2$  & 5/2 &  9161.3~~ &  9186 & -38550  \\
%   &        &                &     & $5d^{9}6s^2$  & 3/2 & 21435.3~~ & 22224 & -26760  \\
%80 & Hg~II  & $5d^{10}6s$    & 1/2 & $5d^{9}6s^2$  & 5/2 & 35514.624 & 35066 & -52200  \\
%   &        &                &     & $5d^{9}6s^2$  & 3/2 & 50555.567 & 50886 & -37700  \\
\hline
\end{tabular}
\end{center}
\end{table*}

It follows from eq.~(\ref{omega}) that the rate at which atomic
frequency and $\alpha$ are changing are related by
\begin{equation}
  \frac{\delta \omega}{\omega} = \frac{2q}{\omega} \frac{\delta \alpha}{\alpha},
\label{enh}
\end{equation}
where $K=2q/\omega$ is an enhancement factor for the relative variation. 
For the most of ``atomic clock transitions''
from the ground state to excited metastable states we have $K < 1$ (see Table ~\ref{tb:opt})
which means no enhancement. However, it is easy to find atomic transitions with
very strong enhancement. One should look for almost degenerate states (small $\omega$)
of sufficiently different nature (large $\Delta q$), e.g. states of different 
configurations. A unique example of this kind is dysprosium atom (see Table~\ref{q-s}).
It has a pair of states of opposite parity which are separated by extremely small
energy interval ($\sim 10^{-4} {\rm cm}^{-1}$). Atomic 
calculations~\cite{Dzuba03c,Dzuba08c} for these states give values of the
$q$ coefficients which are large and have opposite signs. This means that 
the energies of the states move in opposite directions if $\alpha$ is changing
thus adding to the enhancement.

An experiment is currently underway at Berkeley to place limits on
$\alpha$ variation using this transition \cite{budker,budker1}.
The current limit is
 $\dot{\alpha}/\alpha=(-2.7 \pm 2.6) \times 10^{-15}$~yr$^{-1}$.
Unfortunately, one of the levels has  quite a large linewidth
and this limits the accuracy.

Similar pairs of states with strong sensitivity to variation of $\alpha$
can be found in holmium (see Table~\ref{q-s}) and other Rare-Earth 
atoms~\cite{Karshenboim}.

Another way to find large enhancement is to look for fine structure 
anomalies~\cite{Dzuba05c}. Here both states are of the same configuration and
even of the same fine structure multiplet. However strong enhancement 
may exist due to configuration mixing with other states.

\section{Atomic microwave clocks}
Hyperfine microwave transitions may be used to search for $\alpha$-variation \cite{prestage}. 
Karshenboim~\cite{Karshenboim2000} has pointed out that measurements of ratios
of hyperfine structure intervals in different atoms are also sensitive to
variations in nuclear magnetic moments. However, the magnetic moments
are not the fundamental parameters and can not be directly compared with
any theory of the variations. Atomic and nuclear calculations are needed 
for the interpretation of the measurements. We have performed both
atomic calculations of $\alpha$ dependence 
\cite{Dzuba99,Dzuba00c,Dzuba03c,Angstmann,Dzuba05c,Karshenboim,Dzuba08c}
and nuclear calculations of $X_q=m_q/\Lambda_{QCD}$ dependence \cite{tedesco} (see also \cite{thomas})
for all microwave transitions of current experimental interest including
hyperfine transitions in $^{133}$Cs, $^{87}$Rb, $^{171}$Yb$^+$,
$^{199}$Hg$^+$, $^{111}$Cd, $^{129}$Xe, $^{139}$La, $^{1}$H,  $^{2}$H and
$^{3}$He. 

The dependence of the hyperfine transition frequencies on the 
variation of $\alpha$ and fundamental mass ratios can be presented
in the form~\cite{tedesco}
\begin{eqnarray}
  \frac{\delta(A/E)}{A/E} = \frac{\delta V}{V} \nonumber \\
 V = \alpha^{2+K_{rel}}\left(\frac{m_q}{\Lambda_{QCD}}\right)^{\kappa}
\frac{m_e}{m_p},
\label{hfs-al}
\end{eqnarray}
where $A$ is the hyperfine structure constant, $E=m_ee^4/\hbar^2$ is
atomic unit of energy, $m_q$ is quark mass, $\Lambda_{QCD}$ is the quantum 
chromodynamics mass scale, and $K_{rel}$ and $\kappa$ are sensitivity
factors. $K_{rel}$ is the relativistic factor. It is due to the fact
that relativistic hfs depends on $\alpha$ stronger that just $\alpha^2$
as it would be in the non-relativistic limit. The $\kappa$ parameter
consists of at least two contributions, $\kappa=\kappa_{\mu} + \kappa_{hr}$,
where $\kappa_{\mu}$ is due to the dependence of the nuclear magnetic
moments on the mass ratio, and $\kappa_{hr}$ is due to the dependence 
of the hfs constants on nuclear radius (hadron radius factor).
The results of the atomic many-body calculations for $K_{rel}$~\cite{Dzuba99}
and nuclear calculations for $\kappa_{\mu}$~\cite{tedesco} and
$\kappa_{hr}$~\cite{wiringa} are presented in Table~\ref{tab:e}.
One can use these numbers to put limits on the variation of the
fundamental constants which follow from the measurements.
For example, the frequency ratio $Y$ of the 282-nm $^{199}$Hg$^+$ 
optical clock transition to the ground state hyperfine transition
in $^{133}$Cs was monitored to very high precision for over 6 years
at NIST which lead to the following limit on its variation~\cite{Fortier}
\begin{equation}
\dot{Y}/Y=(0.37 \pm 0.39) \times 10^{-15} \ {\rm yr}^{-1}.
\label{YHgCs}
\end{equation}
Using numbers from Tables \ref{tb:opt} and \ref{tab:e} one can get 
the following dependence of this value on the fundamental constants
\begin{equation}\label{Hg}
\dot{Y}/Y=-6\dot{\alpha}/\alpha -\dot{\mu}/\mu +0.02 \dot{X_q}/X_q.
\end{equation}
Here $X_q= m_q/\Lambda_{QCD}$.
For the $\dot{\mu}/\mu$ term we can use the quasar result of Ref.~\cite{FK1}
(see section \ref{ammonia} for a detailed discussion)
\begin{equation}
 \dot{\mu}/\mu=\dot{X_e}/X_e=(1 \pm 3) \times 10^{-16} \ {\rm yr}^{-1}, 
\label{dmu}
\end{equation}
Where $X_e= m_e/\Lambda_{QCD}$.
A combination of this result and the atomic clock result (\ref{YHgCs})
gives one of the best limit on the variation of $\alpha$:
\begin{equation}
\dot{\alpha}/\alpha=(-8 \pm 8) \times 10^{-17} \ {\rm yr}^{-1}.
\label{Hg-alpha}
\end{equation}
Here we neglected the small ($\sim 2\%$) contribution of $X_q$.
%If we use the Hg$^+$-Al$^+$ result (\ref{HgAl}) for the $\dot{\alpha}/\alpha$
%instead then we get
The combined result of Hg$^+$/Al$^+$,  Yb$^+$/Cs$^+$
and  Hg$^+$/Cs$^+$ gives~\cite{Yb06,Fortier,Rosenband}
\begin{equation}
 \dot{\mu}/\mu=\dot{X_e}/X_e=(-0.19 \pm 0.41) \times 10^{-15} \ {\rm yr}^{-1}. 
\end{equation}
which does not contradicts to (\ref{dmu}).

\begin{table*}
\begin{center}
\caption{Sensitivity of the hyperfine transition frequencies to
variation of $\alpha$ (parameter $K_{rel}$) and to
the quark mass/strong interaction scale $m_q/\Lambda_{QCD}$ 
(parameter $\kappa=\kappa_{\mu}+\kappa_{hr}$). 
These values can be used in equation (\ref{hfs-al}).}
\label{tab:e}
%\vspace{2mm}
%\begin{tabular}{@{\extracolsep{3mm}}c|cccccccccc}
\begin{tabular}{ccccccccccc}
\hline
%\hline \\[-2.5mm]
Atom   &  $_1^1$H &  $_1^2$H &  $_{2}^{3}$He    &  $_{37}^{87}$Rb  &  $_{48}^{111}$Cd &  
$_{54}^{129}$Xe &  $_{55}^{133}$Cs & $_{57}^{139}$La & $_{70}^{171}$Yb &  $_{80}^{199}$Hg\\
\\[-3mm] \hline %\\[-2.5mm]
$K_{rel}$   &    0    &   0   &    0    &   0.34     &   0.6   &  0.8  &   0.83    &   0.9  &   1.5   &  2.28\\
\\[-3mm] %\hline \\[-2.5mm]
$\kappa_{\mu}$  & -0.100  & -0.064  & -0.117  & -0.016 &~0.125 &~0.042 &~0.009 & -0.008 & -0.085 & -0.088\\
$\kappa_{hr}$   &   0  &   0     &   0     & -0.010 &-0.018 &-0.024 &-0.025 & -0.028 & -0.051 & -0.081\\
$\kappa$        &-0.100&-0.064  & -0.117  & -0.026 &~0.107 &~0.018 &-0.016 & -0.036 & -0.136 & -0.169\\
\hline
\end{tabular}
\end{center}
\end{table*}

\section{Enhanced effect of variation of fundamental constants
in nuclear spectra ($^{229}$Th)}

There is a very  narrow level  of $7.6\pm 0.5$ eV above the ground state
in the $^{229}$Th nucleus~\cite{th7}. The position
of this level was determined from the energy differences of many high-energy
$\gamma$-transitions to the ground and excited
 states. The subtraction  produces
the large  uncertainty in the position of the 7.6 eV excited state.
 The width of this level is estimated to be
about $10^{-4}$ Hz \cite{th2}. This would explain why it is so hard to find
 the direct radiation in this very weak  transition.
 However, the search for the direct radiation continues
\cite{private}. 

  The  $^{229}$Th transition is very narrow and can be investigated
 with laser spectroscopy.
 This makes $^{229}$Th a possible reference for an
 optical clock of very high accuracy, and opens a new possibility
for a laboratory search for the variation of the fundamental constants
\cite{th4}.

As it is shown in Ref.~\cite{th1} there is an additional very important
 advantage.
According to Ref. \cite{th1} the relative effects of variation of
 $\alpha$ and $m_{q}/\Lambda_{QCD}$ are enhanced by 5 orders of magnitude.
This estimate has been confirmed by the recent calculation \cite{He} and preliminary results of  our
new calculations. The accurate results of the calculations will be published soon.
 A rough estimate for the relative variation of the $^{229}$Th
 transition frequency  is
 \begin{equation}\label{deltaf}
\frac{\delta \omega}{\omega} \approx 10^5 (
0.1 \frac{\delta \alpha}{\alpha} +   \frac{\delta X_q}{X_q })
\end{equation} 
where $X_q=m_q/\Lambda_{QCD}$.
Therefore, the Th  experiment would
have the potential of improving the  sensitivity to temporal
variation of the fundamental
constants by many orders of magnitude. 
Indeed, we obtain the following energy shift
in 7.6 eV $^{229}$Th transition:
\begin{equation}\label{delta3}
\delta \omega \approx 
\frac{\delta X_q}{X_q}  MeV
\end{equation}
This corresponds to the frequency shift
$\delta \nu \approx 3\cdot 10^{20} \delta X_q/X_q$ Hz.
The width of this transition is $10^{-4}$ Hz so one may hope
to get the sensitivity to the variation of $X_q$ about $10^{-24}$
per year. This is  $10^{10}$ times better than the current atomic clock
limit on the variation of $X_q$,  $\sim 10^{-14}$ per year.
 
Note that there are other narrow low-energy levels in nuclei,
 e.g. 76 eV level in $^{235}U$ with the 26.6 minutes lifetime
 (see e.g.\cite{th4}). One may expect a similar  enhancement there.
Unfortunately, this level can not be reached with usual lasers. In principle,
 it may be investigated using a free-electron laser or synchrotron radiation.
However, the accuracy
of the frequency measurements is much lower in this case.

\section{Enhancement of variation of
fundamental constants in atomic and molecular collisions}
Scattering length $A$, which can be measured in Bose-Einstein condensate
and Feshbach molecule experiments, is extremely sensitive to the
variation of the
electron-to-proton mass ratio $\mu=m_e/m_p$ or $X_e=m_e/\Lambda_{QCD}$
 \cite{chin}.
\begin{equation}\label{d_a_final}
\frac{\delta A}{A}=K\frac{\delta\mu}{\mu}=K\frac{\delta X_e}{X_e},
\end{equation}
 where $K$ is the  enhancement factor.
For example, for Cs-Cs collisions we obtained
 $K\sim 400$. With the Feshbach resonance, however, one is
given the flexibility to adjust position of the resonance using
external  fields.  Near a narrow magnetic or an optical Feshbach resonance
 the enhancement factor $K$ may be increased by many orders of magnitude. 

\section{Enhanced relative effect of variation of fundamental constants
in molecular spectra}

A detailed discussion of the variation of fundamental constants in
molecular spectra can be found in recent review~\cite{kozlovmol}.
Below we present several examples.

\subsection{Comparison of hydrogen hyperfine and molecular rotational spectra in quasar data}
\label{rot}
The frequency of the hydrogenic hyperfine line is proportional to
$\alpha^2\mu g_p$ atomic units, molecular rotational frequencies are proportional to 
 $\mu$ atomic units. The comparison places limit on the
variation of the parameter $F=\alpha^2 g_p$~\cite{DWB98}. Recently
similar analysis was repeated by Murphy et al \cite{MWF01d} using
more accurate data for the same object at $z=0.247$ and for a more
distant object at $z=0.6847$, and the following limits were
obtained:
\begin{equation}\label{rotCO2}
 \frac{\delta F}{F} = (-2.0\pm 4.4)\times 10^{-6}
\end{equation}
 \begin{equation}\label{rotCO}
 \frac{\delta F}{F} = (-1.6\pm 5.4)\times 10^{-6}
\end{equation}
The object at $z=0.6847$ is associated with the gravitational lens
toward quasar B0218+357 and corresponds to the backward time $\sim
6.5$ Gyr.

\subsection{Enhancement of variation of $\mu$ in
inversion spectrum of ammonia and limit from quasar data} \label{secNH3}
\label{ammonia}
Few years ago van Veldhoven et al suggested to use decelerated
molecular beam of ND$_3$ to search for the variation of $\mu$ in
laboratory experiments \cite{ammonia}. Ammonia molecule has a
pyramidal shape and the inversion frequency depends on the
exponentially small tunneling of three hydrogens (or deuteriums)
through the potential barrier. Because of that, it is
very sensitive to any changes of the parameters of the system,
particularly to the reduced mass for this vibrational mode. This fact was used in
\cite{FK1} to place the best limit on the cosmological  variation of $\mu$.

The inversion vibrational mode of ammonia is described by a double
well potential with first two vibrational levels lying below the
barrier. Because of the tunneling, these two levels are split in
inversion doublets. The lower doublet corresponds to the wavelength
$\lambda\approx 1.25$~cm and is used in ammonia masers. Molecular
rotation leads to the centrifugal distortion of the potential curve.
Because of that, the inversion splitting depends on the rotational
angular momentum $J$ and its projection on the molecular symmetry
axis $K$:
 \begin{equation}\label{w_inv}
 \omega_\mathrm{inv}(J,K) = \omega^0_\mathrm{inv}
 - c_1
 \left[J(J+1)-K^2\right] + c_2 K^2 + \cdots \,,
 \end{equation}
where we omitted terms with higher powers of $J$ and $K$.
Numerically, $\omega^0_\mathrm{inv}\approx 23.787$~GHz, $c_1\approx
151.3$~MHz, and $c_2\approx 59.7$~MHz.

In addition to the rotational structure (\ref{w_inv}) the inversion
spectrum includes much smaller hyperfine structure. For the main
nitrogen isotope $^{14}$N, the hyperfine structure is dominated by
the electric quadrupole interaction ($\sim 1$~MHz).
Because of the dipole selection rule $\Delta K=0$ the levels with
$J=K$ are metastable. In astrophysics the lines with $J=K$ are
also narrower and stronger than others, but the hyperfine structure
for spectra with high redshifts is still not resolved.
We obtained the following results for NH$_3$ \cite{FK1} (in atomic units):
\begin{equation}
 \label{dw_inv6}
 \frac{\delta\omega_\mathrm{inv}^0}{\omega_\mathrm{inv}^0}
 \approx 4.46\, \frac{\delta\mu}{\mu}\,.
\end{equation}
 \begin{equation}
 \label{dw_rot5}
 \frac{\delta c_{1,2}}{c_{1,2}}
 = 5.1\frac{\delta\mu}{\mu}\,.
 \end{equation}
For ND$_3$ the inversion frequency is 15 times smaller and this
leads to a higher relative sensitivity of the inversion frequency to
$\mu$:
\begin{equation}
 \label{dw_inv6nd3}
 \frac{\delta\omega_\mathrm{inv}^0}{\omega_\mathrm{inv}^0}
 \approx 5.7\, \frac{\delta\mu}{\mu}\,.
\end{equation}
 \begin{equation}
 \label{dw_rot5nd3}
 \frac{\delta c_{1,2}}{c_{1,2}}
 = 6.2\frac{\delta\mu}{\mu}\,.
 \end{equation}
We see  that the
inversion frequency $\omega_\mathrm{inv}^0$ and the rotational
intervals
$\omega_\mathrm{inv}(J_1,K_1)-\omega_\mathrm{inv}(J_2,K_2)$ have
different dependencies on the constant $\mu$. In principle, this
allows one to study time-variation of $\mu$ by comparing different
intervals in the inversion spectrum of ammonia. For example, if we
compare the rotational interval to the inversion frequency, then
Eqs. (\ref{dw_inv6}) and (\ref{dw_rot5}) give:
\begin{equation}
 \label{red1}
 \frac{\delta\{[\omega_\mathrm{inv}(J_1,K_1)-\omega_\mathrm{inv}(J_2,K_2)]
 /\omega^0_\mathrm{inv}\}}
 {[\omega_\mathrm{inv}(J_1,K_1)-\omega_\mathrm{inv}(J_2,K_2)]/\omega^0_\mathrm{inv}}
 = 0.6 \frac{\delta\mu}{\mu}\,.
\end{equation}
The relative effects are substantially larger if we compare the
inversion transitions with the  transitions between the quadrupole
and magnetic hyperfine components. However, in practice this method
will not work because of the smallness of the hyperfine structure
compared to typical line widths in astrophysics.

We  compared the inversion spectrum of NH$_3$ with
rotational spectra of other molecules, where
\begin{equation}
 \label{red2}
 \frac{\delta\omega_\mathrm{rot}}{\omega_\mathrm{rot}}
 = \frac{\delta\mu}{\mu}\,.
\end{equation}
High precision data on the redshifts of NH$_3$ inversion lines exist
for already mentioned object B0218+357 at $z\approx 0.6847$
\cite{HJK05}. Comparing them with the redshifts of rotational lines
of CO, HCO$^+$, and HCN molecules from Ref.~\cite{CW97} one can get
the following  limit:
\begin{equation} \label{nh3final}
 \frac{\delta\mu}{\mu}=\frac{\delta X_e}{X_e}=(-0.6 \pm 1.9)\times 10^{-6}.
\end{equation}
Taking into account that the redshift $z\approx 0.68$ for the object
B0218+357 corresponds to the backward time about 6.5 Gyr and assuming linear
time dependence, this limit
translates into the most stringent present limit 
for the variation rate $\dot\mu/\mu$ and $X_e$ 
\cite{FK1}:
 \begin{equation}\label{best_mu_dot}
 \dot{\mu}/\mu=\dot{X_e}/X_e=(1 \pm 3) \times 10^{-16}\mathrm{~yr}^{-1}\,.
 \end{equation}

\section{Enhanced effects in diatomic molecules}
\label{diatomics}

In transitions between very close narrow levels of different
nature in diatomic molecules the relative effects of the variation
may be enhanced by several orders of magnitude. Such levels may occur due to
cancellation between either hyperfine and rotational structures
\cite{mol}, or between the fine and vibrational structures of the
electronic ground state \cite{FK2}. The intervals between the levels
are conveniently located in microwave frequency range and the level
widths are very small, typically $\sim 10^{-2}$~Hz.

\subsection{Molecules with cancellation between hyperfine
structure and rotational intervals} \label{hfs-rot}

Consider diatomic molecules with unpaired electron and ground state
$^2\Sigma$. It can be, for example, LaS, LaO, LuS, LuO, YbF,
etc.~\cite{HH79}. For a hyperfine interval $\Delta_\mathrm{hfs}$ we
have
\begin{eqnarray}
  \Delta_\mathrm{hfs} \propto \alpha^2 Z F_\mathrm{rel}(\alpha Z) \mu g_\mathrm{nuc}, 
\nonumber
\end{eqnarray}
where $F_\mathrm{rel}$ is additional relativistic
(Casimir) factor.
% \cite{Sob79}.
 Rotational interval
$\Delta_\mathrm{rot} \sim \mu$ is roughly independent on $\alpha$.
If we find molecule with $\Delta_\mathrm{hfs} \approx
\Delta_\mathrm{rot}$ the splitting $\omega$ between hyperfine and
rotational levels will depend on the following combination
\begin{equation}
\label{hfs-rot1}
 \omega \sim  \left[\alpha^2 F_\mathrm{rel}(\alpha Z)\, g_\mathrm{nuc}
 - \mathrm{const}\right]\, .
\end{equation}
Relative variation is then given by
\begin{equation}
\label{hfs-rot2}
 \frac{\delta\omega}{\omega}
 \approx \frac{\Delta_\mathrm{hfs}}{\omega}
 \left[\left(2+K\right)\frac{\delta\alpha}{\alpha} + \frac{\delta
 g_\mathrm{nuc}}{g_\mathrm{nuc}}\right]\,,
\end{equation}
where factor $K$ comes from variation of $F_\mathrm{rel}(\alpha Z)$,
and for $Z \sim 50$, $K\approx 1$.
 Using data from
\cite{HH79} one can find that $\omega = (0.002\pm 0.01)$~cm$^{-1}$
for ${}^{139}$La${}^{32}$S \cite{mol}. Note that for $\omega =
0.002$~cm$^{-1}$ the relative frequency shift is:
\begin{equation}
\label{hfs-rot3}
 \frac{\delta\omega}{\omega}
 \approx 600\,\frac{\delta\alpha}{\alpha}\,.
\end{equation}

% based on file 9b.tex (version 30/08/07 for arXiv) (includes more refs to e-prints)

\subsection{Molecules with cancellation between fine
structure and vibrational intervals} \label{fs-vib}

The fine structure interval $\omega_f$ rapidly grows with nuclear
charge Z:
\begin{equation}
\label{of}
 \omega_f \sim Z^2 \alpha^2\, ,
\end{equation}
The vibration energy quantum decreases with the
atomic mass:
\begin{equation}
 \label{ov}
\omega_\mathrm{vib} \sim M_r^{-1/2} \mu^{1/2}\, ,
\end{equation}
where the reduced mass for the molecular vibration is $M_r m_p$.
Therefore, we obtain equation $Z=Z(M_r,v)$ for the lines on the
plane $Z,M_r$, where we can expect approximate cancellation between
the fine structure and vibrational intervals:
\begin{equation}
 \label{o}
 \omega=\omega_f - v\,  \omega_\mathrm{vib} \approx 0 \,,
 \quad v=1,2,...
\end{equation}
Using Eqs.~(\ref{of}--\ref{o}) it is easy to find dependence of the
transition frequency on the fundamental constants:
\begin{equation}
 \label{do}
 \frac{\delta\omega}{\omega}=
 \frac{1}{\omega}\left(2 \omega_f \frac{\delta\alpha}{\alpha}+
 \frac{v}{2} \omega_\mathrm{vib} \frac{\delta\mu}{\mu}\right)
 \approx K \left(2 \frac{\delta\alpha}{\alpha}+
\frac{1}{2} \frac{\delta\mu}{\mu}\right),
\end{equation}
where the enhancement factor $K= \frac{\omega_f}{\omega}$
%>mgk11/07 and 29/08>
determines the relative frequency shift for the given change of
fundamental constants. Large values of factor $K$ hint at
potentially favorable cases for making experiment, because it is
usually preferable to have larger relative shifts. However, there is
no strict rule that larger $K$ is always better. In some cases, such
as very close levels, this factor may become irrelevant. Thus, it is
also important to consider the absolute values of the shifts and
compare them to the linewidths of the corresponding transitions.

Because the number of molecules is finite we can not have $\omega=0$
exactly. However, a large number of molecules have $\omega/\omega_f
\ll 1$ and $|K| \gg 1$. Moreover, an additional ``fine tuning'' may
be achieved by selection of isotopes and rotational,
$\Omega$-doublet, and hyperfine components. Therefore, we have two
large manifolds, the first one is build on the electron fine
structure excited state and the second one is build on the
vibrational excited state. If these manifolds overlap one may select
two or more transitions with different signs of $\omega$. In this
case expected sign of the $|\omega|$-variation must be different
(since the variation $\delta \omega$ has the same sign) and one can
eliminate some systematic effects. Such control of systematic
effects was used in \cite{budker,budker1} for transitions between
close levels in two dysprosium isotopes. The sign of energy
difference between two levels belonging to different electron
configurations was different in $^{163}$Dy and $^{162}$Dy.

Among the interesting molecules  where the ground state is split in two fine
structure levels and (\ref{o}) is approximately fulfilled, there are
Cl$_2^+$ (enhancement $K=1600$),  SiBr ($K=360$), CuS ($K=24$) and 
 IrC ($K=160$). The list of molecules  is not complete because of the
lack of data in \cite{HH79}. The
molecules Cl$_2^+$ and SiBr are particularly interesting. For both
of them the frequency $\omega$ defined by (\ref{o}) is of the order
of 1~cm$^{-1}$ and comparable to the rotational constant $B$. That
means that $\omega$ can be reduced further by the proper choice of
isotopes, rotational quantum number $J$ and hyperfine components, so
we can expect $K \sim 10^3-10^5$. New dedicated measurements are
needed to determined exact values of the transition frequencies and
find the best transitions. However, it is easy to  find necessary
accuracy of the frequency shift measurements. According to (\ref{do})
the expected frequency shift is
\begin{equation}
\label{do1}
 \delta\omega=2 \omega_f \left(\frac{\delta\alpha}{\alpha}+
 \frac{1}{4}\frac{\delta\mu}{\mu}\right)
\end{equation}
Assuming $\delta \alpha / \alpha \sim 10^{-15}$ and $\omega_f\sim
500$~cm$^{-1}$, we obtain $\delta\omega \sim 10^{-12}$ cm$^{-1}\sim
3 \times 10^{-2}$ Hz  (in order to obtain similar sensitivity
comparing hyperfine transition frequencies for Cs and  Rb one has to
measure the shift $\sim 10^{-5}$ Hz). This shift is larger than the
 natural width $\sim 10^{-2}$ Hz. 

\subsection{Molecular ion {H\lowercase{f}F}$^+$}

 The ion HfF$^+$ and other
similar ions are considered by Cornell's group in JILA for the
experiment to search for the electric dipole moment (EDM) of the
electron. Recent calculation by
\cite{PMI06} suggests that the ground state of this ion is
$^1\Sigma^+$ and the first excited state $^3\Delta_1$ lies only
1633~cm$^{-1}$ higher. Calculated vibrational frequencies for these
two states are 790 and 746~cm$^{-1}$ respectively. For these
parameters the vibrational level $v=3$ of the ground state is only
10~cm$^{-1}$ apart from the $v=1$ level of the state $^3\Delta_1$.
Thus, instead of (\ref{o}) we now have:
\begin{equation}
 \label{hff1}
 \omega=\omega_\mathrm{el} + \frac32 \omega_\mathrm{vib}^{(1)}
 - \frac72\omega_\mathrm{vib}^{(0)}\approx 0\,,
\end{equation}
where superscripts 0 and 1 correspond to the ground and excited
electronic states. Electronic transition $\omega_\mathrm{el}$ is not
a fine structure transition and (\ref{of}) is not applicable.
Instead,  we can write:
\begin{equation}
 \label{hff2}
 \omega_\mathrm{el}=\omega_\mathrm{el,0} + q x\,,
 \quad x=\alpha^2/\alpha_0^2-1\,.
\end{equation}
Our estimate is \cite{FK2}
\begin{equation}
 \label{hff3}
 \frac{\delta\omega}{\omega}
 \approx
 \left(\frac{2q}{\omega} \frac{\delta\alpha}{\alpha}+
 \frac{\omega_\mathrm{el}}{2\omega} \frac{\delta\mu}{\mu}\right)
 \approx \left(2000 \frac{\delta\alpha}{\alpha}+
 80 \frac{\delta\mu}{\mu}\right),
\end{equation}
\begin{equation}
 \label{hff4}
 \delta\omega
 \approx 20000~\mathrm{cm}^{-1}(\delta\alpha/\alpha+0.04 \delta\mu/\mu)\,.
\end{equation}
Assuming $\delta \alpha / \alpha \sim 10^{-15}$ we obtain
$\delta\omega \sim$ 0.6 Hz. The natural width is about 2 Hz.

We also present the result for transition between close levels
in Cs$_2$ molecule suggested in \cite{DeMille,DeMille1}. Our estimate is
\cite{kozlovmol}:
\begin{equation}\label{Cs2c}
 \delta\omega \approx (-240\frac{\delta\alpha}{\alpha}
 -1600\frac{\delta\mu}{\mu}) cm^{-1}\,,
\end{equation}

\section{Change of fundamental constants near massive bodies}

Variation of fundamental constants are related to the change of
the gravitational potential as ~\cite{FS2007}
%$k_i$ as follows
\begin{equation}
 {\delta \alpha \over \alpha } = k_\alpha \delta ({GM\over r c^2})
\label{kalpha}
\end{equation}
\begin{equation}
  {\delta (m_q/\Lambda_{QCD}) \over (m_q/\Lambda_{QCD})} = k_q \delta ({GM\over r c^2})
\label{kmq}
\end{equation}
\begin{equation}
  {\delta (m_e/\Lambda_{QCD}) \over (m_e/\Lambda_{QCD}) } = {\delta (m_e/m_p) 
\over (m_e/m_p) } =k_e \delta ({GM\over r c^2})
\label{kme}
\end{equation}
where in the r.h.s. stands for the half-year variation of Sun's gravitational potential
on Earth. 

Gravitational potential on Earth is changing due to ellipticity of its
orbit,
the corresponding variation of the Sun gravitational potential is
 $\delta (GM/rc^2)=3.3 \cdot 10^{-10}$. 
As an example we consider recent
work~\cite{Fortier} who obtained the following value for the half-year
variation
of the frequency ratio of two atomic clocks: (i) optical transitions in 
mercury ions $^{199}Hg^+$ and (ii) hyperfine splitting
in $^{133}Cs$ (the frequency standard). The limit obtained is
\begin{equation}
 \delta ln({\omega_{Hg}\over \omega_{Cs}})=(0.7\pm 1.2) \cdot 10^{-15}
\end{equation}  
For  Cs/Hg  frequency ratio of these clocks  the dependence on the fundamental
constants
was evaluated in \cite{tedesco} with the result
\begin{equation}
 \delta ln({\omega_{Hg}\over \omega_{Cs}})=-6 {\delta \alpha \over
\alpha} -0.01{\delta  (m_q/\Lambda_{QCD}) 
\over (m_q/\Lambda_{QCD})} -{\delta  (m_e/m_p) \over (m_e/m_p)}
\end{equation}
Another work \cite{BW} compare $H$ and $^{133}Cs$ hyperfine transitions.
The amplitude of the half-year variation  found were
\begin{equation}
 |\delta ln(\omega_{H}/\omega_{Cs})| <7 \cdot 10^{-15}
\end{equation}
The sensitivity \cite{tedesco}
\begin{equation}
\delta ln({\omega_{H}\over \omega_{Cs}})=-0.83 {\delta \alpha \over
\alpha} -0.11{\delta  (m_q/\Lambda_{QCD}) \over (m_q/\Lambda_{QCD})}
\end{equation}
There is no sensitivity to $m_e/m_p$ because
they are both hyperfine transitions.

The results of  Cs/Hg  frequency ratio measurement
\cite{Fortier}  can be rewritten in terms of the parameters $k_i$ as
in (\ref{kalpha},\ref{kmq},\ref{kme}):
\begin{equation}
%% k_\alpha +0.17 k_e= (-3.5\pm 6) \cdot 10^{-7}
 k_\alpha +0.17 k_e= (3.5\pm 6.0) \cdot 10^{-7}
\end{equation}
 The results of  Cs/H  frequency ratio measurement \cite{BW} can be presented as 
\begin{equation}
| k_\alpha +  0.13 k_q | <2.5 \cdot 10^{-5}
\end{equation}
Finally, the result of recent measurement  \cite{Ashby} of
  Cs/H  frequency ratio can be presented as 
\begin{equation}
%% k_\alpha +0.13 k_q= (-1\pm 17) \cdot 10^{-7}
 k_\alpha +0.13 k_q= (1\pm 17) \cdot 10^{-7}
\end{equation}
The sensitivity coefficients for other  clocks have been discussed above.

Two new results have been obtained recently. From transition between
close levels in Dy we obtain \cite{budkerG}
\begin{equation}
%% Should this be corrected as well???????????
 k_\alpha = (-8.7\pm 6.6) \cdot 10^{-6}
\end{equation}
From optical  Sr/hyperfine Cs comparison we obtained \cite{Sr08}
\begin{equation}
%% k_\alpha+0.36 k_e = (1.8\pm 3.2) \cdot 10^{-6}
 k_\alpha+0.36 k_e = (-2.1\pm 3.2) \cdot 10^{-6}
\end{equation}
Combination of the data gives \cite{Sr08}
\begin{equation}
%% k_\alpha = (-2.3\pm 3.1) \cdot 10^{-6}
 k_\alpha = (2.5\pm 3.1) \cdot 10^{-6}
\end{equation}
\begin{equation}
%% k_e = (1.1\pm 1.7) \cdot 10^{-5}
 k_e = (-1.3\pm 1.7) \cdot 10^{-5}
\end{equation}
\begin{equation}
%% k_\alpha = (1.7\pm 2.7) \cdot 10^{-5}
 k_q = (-1.9\pm 2.7) \cdot 10^{-5}
\end{equation}

\section*{Acknowledgment}

We are grateful to Sebastian Blatt who brought to our attention 
sign errors in cited papers.
The work was supported in part by the Australia Research Council.

\end{document}